# Iterative Random Weight EC-TOPSIS Method for Ranking Companies on Social Media

**Marcio Pereira Basilio**

**marciopbasilio@gmail.com**

**Department of Production Engineering, Federal Fluminense University, RJ, Brazil.**

**ORCID: 0000-0002-9453-741X**

**Valdecy Pereira**

**valdecy.pereira@gmail.com**

**Department of Production Engineering, Federal Fluminense University, RJ, Brazil.**

**ORCID: 0000-0003-0599-8888**

**Fatih Yigit**

**fatih.yigit@altinbas.edu.tr**

**Department of Industrial Engineering, Altinbas University, Turkey.**

**ORCID: 0000-0002-7919-544X**

**Büşra Ayan**

**ayanbu@mef.edu.tr**

**Business Administration Department, MEF University, Istanbul, Turkey**

**ORCID: 0000-0002-5212-2144**

**Seda Abacıoğlu**

**seda.abacioglu@omu.edu.tr**

**Department of Business Administration, Ondokuz Mayıs University, Samsun, Turkey**

**ORCID: 0000-0002-3150-2770**

**Abstract:** This study aims to present a new hybrid method for weighting criteria. The methodological project combines the ENTROPY and CRITIC methods with the TOPSIS method to create EC-TOPSIS. The difference lies in the use of a weight range per criterion. Each weight range has a lower limit and an upper limit, which are combined to generate random numbers, producing "t" sets of weights per criterion, allowing "t" final rankings to be obtained. The final ranking is obtained by applying the MODE statistical measure to the set of "t" positions of each alternative. The method was validated by ranking the companies based on social media metrics consisting of user-generated content (UGC). The result was compared with the original modeling using the CRITIC-ARAS and CRITIC-COPRAS methods, and the results were consistent and balanced, with few changes. The practical implication of the method is in reducing the uncertainties surrounding the final classification due to the random weighting process and the number of interactions sent.

## 1. Introduction

The Technique for Order Performance by Similarity to Ideal Solution (TOPSIS), developed by Hwang and Yoon (Hwang and Yoon, 1981) and after forty-three years, with more than twenty-one papers published on the Scopus database, the TOPSIS method is the second most used method by experts according to Basilio et al. (Basílio *et al.*, 2022). TOPSIS has been applied in numerous areas of knowledge such as: Engineering, Computer Science, Mathematics, Environmental Science, Energy, Business, Social Sciences, Decision Sciences, Astronomy, Agricultural, Biological Sciences, and Medicine. In the evolution of classical methods, we can see that experts have developed numerous hybrid methods in order to improve support for decision-making. In the specialized literature we have identified numerous hybrid models that use the TOPSIS method in its classic alternative ranking function. However, the branch that interests us is the one that combines the methods for objectively obtaining criteria weights with the TOPSIS method. In this sense, we have identified the following methods: AHP-TOPSIS(Dağdeviren, Yavuz and Kılınç, 2009); BWM-TOPSIS (You *et al.*, 2017); SWARA-TOPSIS (Akcan and Taş, 2019); GRA-TOPSIS (Sabry, El-Attar and Hewidy, 2024); CILOS-TOPSIS (Al-Khulaidi *et al.*, 2024); IDOCRIW-TOPSIS (Alao, Popoola and Ayodele, 2021); FUCOM-TOPSIS (Majumder, 2023); MEREC-TOPSIS (Yadav *et al.*, 2023); Entropy-TOPSIS (Liu *et al.*, 2019); CRITIC-TOPSIS (Babatunde and Ighravwe, 2019). What characterizes these methods is simply obtaining the criteria weights and inserting them into the TOPSIS method for ranking purposes. However, other models integrate more than one method for obtaining the weights, such as the following: AHP-Entropy-TOPSIS (′Jalalifar, ′Behaadini and ′Aghajani Bazzazi, 2009; Freeman and Chen, 2015; Zhai, 2024); AHP-CRITIC-TOPSIS (Li *et al.*, 2021; Yu *et al.*, 2024); BWM-CRITIC-TOPSIS (Stark, Wan and Chin, 2022)– in these methods, subjective and objective weights are integrated using the product method; AHP-CILOS-TOPSIS (Tajik, Makui and Tosarkani, 2023)- In this method, subjective and objective weights were combined using the geometric mean; BWM-Entropy-TOPSIS (Liu *et al.*, 2020) – In this method, the subjective and objective weights are merged using a coefficient of participation of the subjective weight about the objective weight.

The space created between experts who advocate using objective methods for obtaining criteria weights removes human intervention from the process, thus trying to eliminate the subjectivity behind the definition of criteria weights. Some advocate only human intervention in this process, as it brings the experience of specialists and intrinsic knowledge of the problems to be solved into the process. However, it can also bring negative subjectivities such as private, corporate, and political interests favouring a particular solution. Experts may find it challenging to arrive at the membership values in the real world objectively (Aggarwal, 2017). Due to their intrinsic dependence on individual experiences and subjective assessments of decision-makers, subjective weight coefficients are prone to individual variances. Although expert opinions are usually used to determine these factors, making decisions based only on subjective evaluations might lead to biases and mistakes. On the other hand, objective approaches ignore ambiguities and inconsistencies in decision-maker assessments by using mathematical models and data from the decision matrix to calculate criteria weights (Paksoy, 2017; Demir, Özyalçın and Bircan, 2021). As a result, a third branch of this discussion emerged: hybrid methods, which seek a balance by combining the weights of criteria obtained by objective and subjective methods. As we have seen in the literature, there is a gap in the integration of criteria weighting methods. In this sense, this research proposes to develop a hybrid method that integrates the criteria weighting methods efficiently. In the seminal work by Basilio et al. (Basilio, Pereira and Yigit, 2023; Basilio, Pereira and Yiğit, 2024), the first method was developed called the EC-PROMETHEE method, which integrated the objective ENTROPY and CRITIC methods with the multi-criteria PROMETHEE method. This work will use the Technique for Order of Preference by Similarity to Ideal Solution (TOPSIS) method. This method was chosen because it is the second most used method by experts worldwide (Basílio *et al.*, 2022). The new method will be called EC-TOPSIS and combines objective and subjective methods for obtaining the weight of the criteria and the possibility of inserting a third external method, which can be objective or subjective. The innovation contained in this method lies in the creation of a weight range for each criterion, preserving the characteristics of each technique. This technique differs from other hybrid methods in that it is not an algebraic combination of the different methods used. In this sense, each weight range comprises lower and upper limits, which can be combined to generate random numbers, producing "t" sets of weights per criterion, making obtaining "t" final classifications possible. The alternatives receive a value corresponding to their position in each ranking generated. At the end of the process, they will be ranked in descending order, thus obtaining the final definitive ranking. In this way, managers can analyse the behaviour of each alternative throughout the process, and the final ranking will be more consistent due to the incorporation of the variations caused by the influence of the weight of the criteria on the alternatives. To apply EC-TOPSIS, the researchers developed an application in Python and made it available at the following address: https://pypi.org/project/ec-topsis/.

To validate the proposed method, we used the results of the research presented by Ayan and Abacıoğlu(Ayan and Abacıoğlu, 2022), in which they proposed an MCDM approach to evaluating companies' social media metrics based on UGC (User-Generated Content). In this approach, the researchers used the CRITIC method to obtain the criteria weights and ARAS and COPRAS to rank the companies. The motivation for selecting this topic was due to the possibility of applying multi-criteria decision-making (MCDM) techniques to classify social media companies, given the variety of metrics and criteria that must be considered simultaneously. Although there are some ranking platforms, they are generally not transparent about the calculation process and the importance of the relevant metrics (criteria weights). In addition, there are few studies on social media-based company rankings (Capatina *et al.*, 2018; Irfan *et al.*, 2018), and MCDM methods have not been applied in these studies. The study to which the dataset belongs aims to address this gap by proposing an MCDM approach to determine the weights of platform X (Twitter) metrics and rank companies based on brand-related user-generated content (UGC). The results using EC-TOPSIS were consistent. In addition to validation, we compared the results of EC-TOPSIS with the IDOCRIW and MEREC-based ranking methods: ARAS, COPRAS, EDAS, MARCOS, and TOPSIS. This step contributed to proving the robustness of EC-TOPSIS to other methods used by experts.

This article is structured in six different sections. The first is called Background, where we introduce the main concepts used throughout the text. In the second, we present a literature review on the methods used to model EC-TOPSIS. In the third section, we present the proposed method in detail. In the fourth section, we validate the method. In the fifth section, we present the results followed by the discussion. Finally, in the sixth section, we present the conclusions about the prosecution and consistency of the new method.

## 2. Literature Review

In the realm of decision-making and problem-solving, researchers have long sought to develop robust and comprehensive methodologies that can effectively address the complexities inherent in real-world scenarios. Different methods and hybrid approaches are applied to address this issue. The first approach uses various methods, namely AHP, TOPSIS, SWARA, ELECTRE, ANP, and other known methods and their extensions, such as fuzzy logic. The second approach is to use hybrid approaches using multiple methods simultaneously, in succession, including qualitative and quantitative methods. The techniques are used to identify evaluation indicators and their weights. Two primary categories of criteria were used for assessment: subjective and objective (Wei *et al.*, 2024). Liu (Liu *et al.*, 2024) stated that the most critical aspects of Multi-Criteria Decision Analysis (MCDA) lie in selecting indicators and calculating weights. According to a recent study, weight determination using only a single approach has shortcomings; as a result, hybrid approaches have advantages compared to single methods for weight assessment(Wei *et al.*, 2024). A novel hybrid approach is used for the proposed model. In the following sub-sections, the literature review associated with the mentioned methods, the hybrid approaches, and the studies associated with relevant applications will be given.

### 2.1 Entropy, CRITIC and TOPSIS Methods

The following sub-chapters will give a literature review regarding the methods used as part of the model.

#### 2.1.1 Entropy Weight Method (EWM)

The Entropy method is a powerful tool for assessing the relative importance of various criteria in decision-making (Sampathkumar *et al.*, 2023; Altıntaş, 2024). The entropy method is an approach that uses the original data for the alternatives. The other approaches are subjective in one way or another. Also, the importance of using weighting schemes is shown in previous research (Fox *et al.*, 2020). It provides a quantitative measure of the uncertainty or randomness associated with a given data set, allowing decision-makers to determine the degree of information contained within each criterion. The Entropy Weight Method (EWM) is based on Shannon entropy, originally developed by Shannon (Shannon, 1948). Shannon entropy is a concept proposed as a measure of uncertainty in information, formulated in terms of probability theory. The concept of entropy is well suited to measure the relative intensities of contrast criteria in order to represent the average intrinsic information transmitted for decision-making (Zeleny, 1996). This application eliminates biases inherent in subjective approaches. Entropy is particularly effective in identifying influential criteria in environmental management, where variables such as pollutant levels vary significantly. The method uses a degree of uncertainty in data developed concerning probability theory. The relative contrast intensities of the criteria are assessed using the entropy approach. The significant advantage of the Entropy method over other subjective weighting models is the elimination of human involvement with the weight of indicators, which improves the impartiality of the results of the thorough review. The Entropy method is sensitive to the diversity of a criterion; as a result, it performs well in distinguishing alternatives and determines the importance weights of the criteria according to the diversity they capture (P. Chen, 2021; Mete *et al.*, 2023). Recent studies show that Entropy-based weighting has been described as reliable and objective (Joshi and Kumar, 2022; Yue *et al.*, 2024). Since its introduction, it has been widely used for different application areas, for the analysis of wastewater microalgae culture systems for bioenergy production, evaluation of investment environment, selection of arc welding robots, and social problems involving full employment and quality of the employment (Chodha *et al.*, 2022; Zhang, Feng and Feng, 2023; Liu *et al.*, 2024; Wei *et al.*, 2024).

Entropy has been integrated with other methods to enhance decision-making frameworks in recent applications. For instance, Sitorus and Brito-Parada (Sitorus and Brito-Parada, 2020) used Entropy with fuzzy logic to prioritize renewable energy projects under uncertain conditions. Similarly, the hybrid approach uses Entropy in supply chain evaluations, demonstrating its robustness in scenarios with incomplete or imprecise data (dos Santos, Godoy and Campos, 2019). The flexibility that is a part of the hybrid approach motivated the researchers to use it as a part of the proposed study. These studies underscore the method's adaptability and relevance in tackling real-world problems involving complex, multidimensional datasets.

#### 2.1.2 CRITIC Method

The CRITIC method is a technique that determines the objective weights of criteria based on the degree of conflict and correlation between them. CRITIC is a type of correlation method. When selecting the weight, the conflict and contrast intensity between indications must be considered, in addition to avoiding the influence of subjective elements (Liu *et al.*, 2024). The CRITIC considers the conflict between criteria and variation within each criterion to provide more accurate criteria weights. This ability underlines the importance of the CRITIC method as an MCDM method (Chatterjee and Chakraborty, 2024). By analyzing the contrast intensity and the conflicting information in the criteria, the CRITIC method can help decision-makers identify the most influential factors in the decision-making process. The CRITIC method has emerged as a robust and versatile tool for multi-criteria decision-making in various domains. In recent years, the CRITIC method has gained significant attention from researchers and practitioners alike, with a growing body of literature exploring its applications across various fields. The method's ability to effectively capture the conflicting criteria and their relative importance has made it a valuable asset in decision-making processes, particularly in complex scenarios where multiple, often competing, objectives must be considered simultaneously. The CRITIC method avoids the interference of subjective factors and considers the contrast intensity and conflict between indicators to determine the weight. This ability increases the importance of the method (Anwar, Rasul and Ashwath, 2019; Liu *et al.*, 2024).

The existing literature has highlighted the CRITIC method's remarkable versatility and capacity for adaptation(Wang *et al.*, 2021). For example, the entropy weight method could not consider the horizontal influence among evaluation indicators, and the AHP is highly subjective. Kalvakolanu et al. (Kalvakolanu *et al.*, 2022) found a highly significant mutually reinforcing relationship between the weights obtained through Entropy and the CRITIC method (Diakoulaki, Mavrotas and Papayannakis, 1995). The CRITIC method has also found applications in the business and financial sectors, where it has been utilized to support decision-making processes in areas such as investment portfolio optimization and risk management (Behzadian *et al.*, 2010). Moreover, recent research has further explored integrating the CRITIC method with other multi-criteria decision-making techniques, such as the PROMETHEE method. The goal is to enhance the decision-making process's robustness and reliability (Behzadian *et al.*, 2010). The CRITIC method is more scientific and stable (Zhang, Lv and Yuan, 2023). Due to these advantages, many studies have used the CRITIC method to evaluate decision-making problems (Diakoulaki, Mavrotas and Papayannakis, 1995). The empowerment of the Entropy weight method and CRITIC allows us to quickly identify the key factors that could increase the bioenergy production of microalgae culture systems (Liu *et al.*, 2024). The empowerment also motivated the researchers to choose the two models for hybrid application. Also, it was found that there is a highly significant mutually reinforcing relationship between the weights obtained through Entropy and the CRITIC method (Kalvakolanu *et al.*, 2022). Compared with objective weighting methods, subjective methods have the disadvantages of being highly subjective and over-dependence on scoring experts. Therefore, a recent paper uses objective weights based on combining CRITIC and Entropy methods (Wei *et al.*, 2024).

The recent literature on the CRITIC method underscores its growing importance and widespread adoption in various decision-making contexts. The method's ability to effectively incorporate multiple criteria and their relative importance has positioned it as a valuable tool for researchers and practitioners seeking to navigate complex decision-making challenges in a wide range of industries and applications(Steele *et al.*, 2009; Faizi *et al.*, 2018; Makan and Fadili, 2020).

**2.1.3 TOPSIS Method**

TOPSIS is a widely utilized multi-criteria decision-making method that has garnered significant attention. TOPSIS has been applied in various industrial applications, including evaluating suppliers, processes, designs, and even standards (Yadav, Joseph and Jigeesh, 2018). The strategy depends on choosing the option closest to the ideal solution and the furthest from the perfect negative answer (Sari *et al.*, 2018).

The unique approach of TOPSIS, which seeks to identify the optimal solution through proximity to the ideal, has made it a preferred tool for decision-makers in various sectors. For instance, in the chicken slaughterhouse industry, TOPSIS has been employed to select the most suitable raw material suppliers. Similarly, in personnel selection, the method has been adapted to incorporate the concept of a veto threshold, a critical characteristic of outranking methods, to support the decision-making process (Kelemenis and Askounis, 2010). TOPSIS has also found applications in the field of marketing strategy, where it has been combined with the Analytic Network Process to identify the optimal marketing strategy (Wu, Lin and Lee, 2010). The versatility of TOPSIS is further demonstrated by its use in evaluating various entities, such as suppliers, processes, designs, and locations, across a diverse range of industries, including automobile, mobile, information technology, and manufacturing (Yadav, Joseph and Jigeesh, 2018).

The growing popularity of TOPSIS can be attributed to its ease of use and the intuitive nature of its underlying principles. The method's ability to handle both quantitative and qualitative criteria and its robustness in handling uncertainty have made it a valuable tool in the industrial engineering domain.

**2.1.4 Hybrid Approaches**

Multi-criteria decision-making has emerged as a prominent field in decision analysis, providing a structured framework for evaluating and prioritizing alternatives based on multiple, often conflicting, criteria(Faizi *et al.*, 2018; Tian *et al.*, 2018; Conejero *et al.*, 2021). The integration of various MCDM techniques, known as hybrid approaches, has gained significant attention in recent years due to their ability to leverage the strengths of individual methods and address their limitations (Tian *et al.*, 2018). Several MCDM techniques have been independently combined with fuzzy sets, rough numbers, and fuzzy rough numbers to address complex issues (Chakraborty *et al.*, 2023). A recent study presents a hybrid approach that combines the cloud model, CRITIC method, and Probabilistic Dominance Relation (PDR) to address the issues of insufficient uncertainty information measure, inaccurate weight calculation, and incommensurability of indices in hybrid multi-criteria decision-making (Zhang, Feng and Feng, 2023). A method to integrate human experience and judgment is to use fuzzy logic. Nonetheless, the constraint of fuzzy sets is that they solely represent data ambiguity and cannot encapsulate randomness (Zhang, Lv and Yuan, 2023). The advantages of these combined weight assignment methods are that they assign weights by examining experts' opinions and the objectivity of the decision problem (Deepa *et al.*, 2019). The inability to cover randomness by other methods encourages the researchers to propose a method to overcome this limitation.

One such hybrid approach, combining the ELECTRE, CRITIC, and TOPSIS methods, has been the subject of limited investigation, particularly in randomized applications. The ELECTRE method is known for its ability to handle complex decision problems with conflicting criteria. In contrast, the CRITIC method provides a system-atic approach to determining the relative importance of criteria (Wang, Wang and Liu, 2018) The TOPSIS method, on the other hand, offers an intuitive and widely used technique for ranking alternatives based on their proximity to the ideal solution (T.-Y. Chen, 2021).

Integrating these three methods can potentially result in a more comprehensive and robust decision-making process, as it allows for considering multiple perspectives and incorporating both objective and subjective evaluations. Despite the potential benefits of this hybrid approach, the literature review for this study reveals a dearth of research exploring its randomized applications. Emerging trends in hybrid approaches focus on incorporating machine learning and artificial intelligence to automate weight determination and enhance decision accuracy. Integrating Entropy-CRITIC with neural networks enables real-time decision support in dynamic environments (Zhang, Lv and Yuan, 2023). While Entropy, CRITIC, and TOPSIS are powerful individually, their hybridization addresses inherent limitations. For instance, while Entropy emphasizes data dispersion, it does not account for interdependencies among criteria, which CRITIC effectively captures. Integrating these weighting methods with TOPSIS provides a balanced and actionable ranking system. In multi-criteria decision-making, integrating complementary weighting and ranking methods can address inherent limitations and provide a more comprehensive and robust decision-support framework. While individual techniques like Entropy, CRITIC, and TOPSIS possess unique strengths, their hybridization can leverage the advantages of each approach to produce a superior decision-making solution(Frazão *et al.*, 2018; Garg and Kaur, 2018). As a result, the second widely chosen method is a hybrid approach for MCDM. The development of hybrid and modular methods is becoming increasingly important. They are based on previously developed, well-known methods (Mardani, Jusoh and Zavadskas, 2015). Several shortcomings of usual classical MCDM methods can be solved by using the proposed variety of hybrid methods (Zavadskas, Kalibatas and Kalibatiene, 2016)**.** Selecting an appropriate method is a continuous challenge in every situation that requires a decision. Different MCDM methods sometimes yield different rankings of alternatives. No method can be considered best for a general or a particular problem (Saaty and Ergu, 2015). Accordingly, more than one MCDM method was used, and results were integrated for final decision-making (Zavadskas, Kalibatas and Kalibatiene, 2016).

Most existing studies have focused on deterministic scenarios with well-defined input data and parameters. However, in real-world decision-making contexts, there is often a significant degree of uncertainty and randomness involved, which can impact the decision-making process and the reliability of the results. Limitations with combining different methods are used to deal with uncertainty resulting from subjective judgments and vague linguistic evaluations (Görener *et al.*, 2017). Because of accuracy, reliability, flexibility for usage in different contexts, and ease of implementation, multiple methods are widely used, and combining them could be advantageous for decision-makers (Jafari *et al.*, 2020). Integrating two or more methods by utilizing each other's advantages, such as one approach to prioritize decision-making attributes while other approaches are used to find outranking relations (Sitorus and Brito-Parada, 2020). Both subjective and objective weights play crucial roles in determining weighting criteria in methodologies. However, relying solely on subjective weights may result in time-consuming and less accurate assessments, particularly in cases of disagreement among decision-makers. Conversely, considering only objective weights might (Kumar and Mahanta, 2024).

### 2.2 Application Areas

The proposed study proposes a new model and application in areas that are not fully applied. Literature reviews regarding the application areas of policing and social media are given in the following sub-chapters.

#### 2.2.1 Policing

The adoption of community policing strategies in rural areas is an important area of interest, as researchers have noted that small and rural agencies have gradually followed the lead of their larger counterparts. However, the existing knowledge on the amenability of community policing in rural settings remains limited (Pelfrey, 2007). The existing literature suggests that policing in rural communities can present unique challenges for officers, such as blurred professional boundaries in small towns (Huey and Ricciardelli, 2015). Understanding the complexities of rural policing is crucial, as these areas represent a significant portion of the population in many countries. Researchers have developed Decision Support Systems to improve police patrolling, incorporating forecasting and districting models (Camacho-Collados and Liberatore, 2015; Basilio *et al.*, 2024). These systems have demonstrated enhanced performance compared to traditional patrolling methods.

Additionally, studies have examined the effectiveness of patrolling (Basilio *et al.*, 2024), classifying policing strategies (Basilio, Pereira and Brum, 2019; Basilio and Pereira, 2020; Basilio, Brum and Pereira, 2020; Basilio *et al.*, 2020), evaluating target programs(Basilio, Pereira and Costa, 2019) and signaling schemes in preventing poaching, considering real-world constraints like limited resources and dynamic poacher behavior (Hasan *et al.*, 2022). Investigations into the impact of highway patrol efforts on fatality rates have found a negative correlation between enforcement resources and fatality rates (Rezapour, Wulff and Ksaibati, 2018). Furthermore, researchers have evaluated police service quality in handling traffic crash reporting, identifying strengths and weaknesses and proposing improvements to enhance reporting rates (Janstrup *et al.*, 2017).

These studies show that effective policing strategies are vital for the community. MCDM methods are effective solutions to solve policing problems. As a result, the proposed method is applied in a policing area.

#### 2.2.2 Social Media

Social media's influence on customer trust in financial services, mainly through online reviews, is a growing study area (Nalluri and Chen, 2023). Researchers use techniques like NLP and TOPSIS to analyze online reviews and identify factors impacting trust, such as security concerns and firm credibility (Nalluri and Chen, 2023). Concurrently, the success of social media marketing strategies hinges on factors such as content, communication, and security (Jami Pour, Hosseinzadeh and Amoozad Mahdiraji, 2021). Finally, assessing the ability of social media accounts to refute health rumors requires a multi-faceted approach, incorporating social network analysis and evaluation indicators (Yin *et al.*, 2024).

A recent study showed that the in 32 areas, but the majority of research is found in seven areas, including Computer Science, Engineering, Operational Research & Management Science, Business Economics, Mathematics, Energy Fuels, and Environmental Sciences Ecology (Zavadskas, Kalibatas and Kalibatiene, 2016). The study did not mention the application of social media or policing. To the best of our research, the hybrid application of EC and TOPSIS using a random weight assignment within limits is new in the application area of policing and social media. Based on our detailed literature review, the proposed study fills an essential scientific community gap by employing a hybrid approach with critical practical applications.

## 3. Materials and Methods

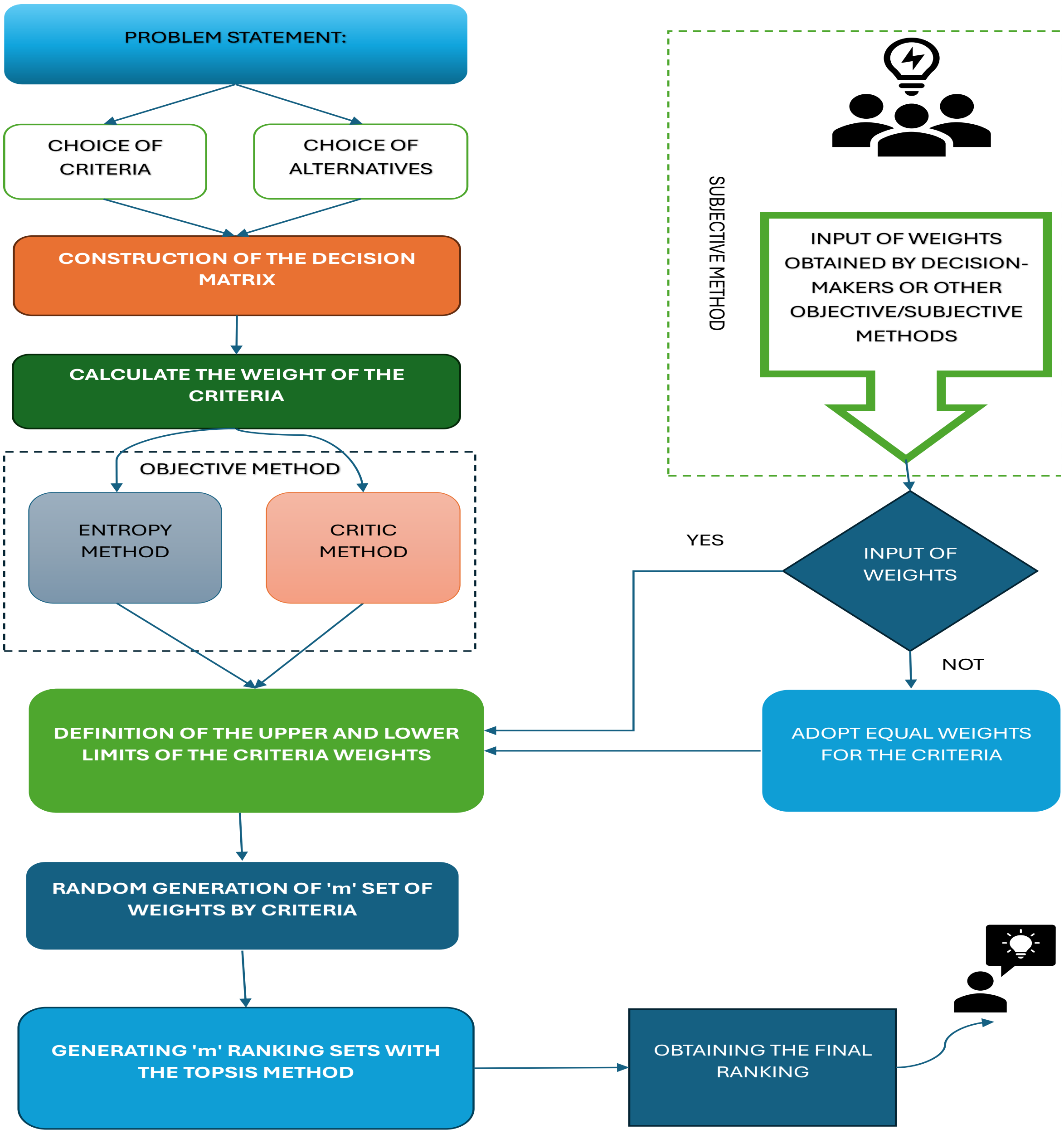

**Figure 1 Methodological scheme.**

This section presents the concepts and formulations for formulating the hybrid EC-TOPSIS method. Figure 1 illustrates the description of the proposed method by subdividing it into eight steps. Figure 1 can be applied to any type of problem in which the decision maker needs to rank alternatives, i.e., a generic problem-solving scheme.

**Step 1 – Identification of criteria**

At this stage, the decision-makers and/or analysts of the problem under study identify the criteria $(c_j)$, belonging to $\mathbf{C}, \mathbf{A}$, $c_j \in \mathbf{C}$, $\mathbf{j} = \mathbf{1}, \mathbf{2}, \mathbf{3}, \ldots, \mathbf{n}$, that will be included in the model to solve the problem.

**Step 2 – Identification of alternative**

At this stage, the decision-makers and/or analysts of the problem under study identify the alternatives$(a_i)$, belonging to a set A, $a_i \in A$, $i = 1, 2, 3, \ldots, m$, that will be included in the model to solve the problem.

**Step 3 – Construction of the decision matrix**

At this stage, the decision-makers and/or analysts of the problem under study collect the information corresponding to the set of alternatives $(a_i)$, belonging to a set A, $a_i \in A$, $i = 1, 2, 3, \ldots, m$, evaluated in the light of a set of criteria $(c_j)$, belonging to $\mathbf{C}, \mathbf{A}$, $c_j \in \mathbf{C}$, $\mathbf{j} = \mathbf{1}, \mathbf{2}, \mathbf{3}, \ldots, \mathbf{n}$, which will be inserted into the model to solve the problem. The Decision Matrix M, in which $a_{ij}$ quantitatively represents the relationship between alternative $a_i$ and criterion $c_j$, can be constructed according to the model illustrated in Table 1.

*Table 1 Generic decision matrix*

| A \ C | $c_1$ | $c_2$ | $c_3$ | … | $c_n$ |
|---|---|---|---|---|---|
| $a_1$ | $a_{11}$ | $a_{12}$ | $a_{13}$ | … | $a_{1n}$ |
| $a_2$ | $a_{21}$ | $a_{22}$ | $a_{23}$ | … | $a_{2n}$ |
| $a_3$ | $a_{31}$ | $a_{32}$ | $a_{33}$ | … | $a_{3n}$ |
| … | … | … | … | … | … |
| $a_m$ | $a_{m1}$ | $a_{m2}$ | $a_{m3}$ | … | $a_{mn}$ |

**Step 4 Description of how the criteria weights are obtained**

**Step 4.1 The ENTROPY WEIGTH METHOD (EWM)**

The criteria weights are based on a predefined decision matrix (DM) comprising information for the set of candidate materials when the Entropy method is used. Entropy in information theory is a model for the uncertainty volume served by a discrete probability distribution (Lau *et al.*, 2018; Mahajan *et al.*, 2022). Salwa et al. (Salwa *et al.*, 2020) used the entropy method to calculate criterion weight to select optimal starch as the matrix in green composites for single-use food packaging applications [21]. The Entropy of the normalized decision matrix (NDM) criterion is given in Eq. (1) as (Zhu, Tian and Yan, 2020):

$$E_j = -\frac{[\sum_{i=1}^{m} P_{ij} \ln(P_{ij})]}{\ln(m)}; j = 1, 2, \ldots, n \text{ and } i = 1, 2, \ldots, m \quad (1)$$

where $P_{ij}$ is NDM, which is given by Eq. (2):

$$P_{ij} = \frac{x_{ij}}{\sum_{i=1}^{m} x_{ij}}; j = 1, 2, \dots, n \; and \; i = 1, 2, \dots, m \quad (2)$$

where $x_{ij}$ corresponds to the criteria value for each alternative in DM. The criteria weight, $W_{1xj}^{E}$ can be calculated using Eq. (3):

$$W_{1xj}^{E} = \frac{1-E_j}{\sum_{j=1}^{n} 1-E_j}; j = 1, 2, \dots, n \quad (3)$$

where $(1 - E_j)$ denotes the degree of diversity of the information in the jth criterion outcome.

**Step 4.2 The CRITIC method**

In this section, a brief description of the CRITIC method is presented. The CRITIC method proposed by (Lau *et al.*, 2018)aims to determine the criteria weights. In this method, the qualitative attributes are replaced with some quantities, and the independence of the attributes is not obligatory. The main steps of this technique can be described as follows:

**Step 4.2.1.** A decision matrix, Z, with $m$ rows as the number of alternatives and $n$ column as the number of criteria, is defined by Eq.(4):

$$Z = (r_{ij})_{mxn}; \; i = 1, \dots, m; j = 1, \dots, n \quad (4)$$

where $r_{ij}$ is the correlation of the ith alternative and the jth criterion.

Step 4.2.2. Each criterion can be considered beneficial or non-beneficial (Wu, Zhen and Zhang, 2020). A criterion takes value in some bounded range. For a beneficial one,$j \in F^{+}$, the criterion is normalized by dividing its distance from the minimum value by the length of the range. In contrast, a non-beneficial one, $\in F^{-}$, is normalized by dividing its distance from the maximum value by the length of the range. The elements of the decision matrix are normalized as given in Eq.(5-6) for the positive or beneficial criteria and the negative or non-beneficial ones.

$$x_{ij}^{+} = \frac{r_{ij} - r_j^{-}}{r_j^{+} - r_j^{-}}; \; i = 1, \dots, m; j = 1, \dots, n \; if \; j \in F^{+} \quad (5)$$

$$x_{ij}^{-} = \frac{r_j^{+} - r_{ij}}{r_j^{+} - r_j^{-}}; \; i = 1, \dots, m; j = 1, \dots, n \; if \; j \in F^{-} \quad (6)$$

where $r_j^{+} = \max(r_{1j}, r_{2j}, \dots, r_{mj})$ and $r_j^{-} = \min(r_{1j}, r_{2j}, \dots, r_{mj})$, and $x_{ij}$ which is either $x_j^{+}$ or $x_j^{-}$represents the normalized value of the $ij$ element of the decision matrix.

Step 4.2.3. The Pearson correlation coefficient between two criteria, j and k, is computed as Eq. (7)

$$\rho_{jk} = \frac{\sum_{i=1}^{m} (x_{ij} - \underline{x_j})(x_{ik} - \underline{x_k})}{\sqrt{\sum_{i=1}^{m} (x_{ij} - \underline{x_j})^2 \sum_{i=1}^{m} (x_{ik} - \underline{x_k})^2}} \quad (7)$$

where $\underline{x_j}$ and $\underline{x_k}$ represent the mean of *jth* and *kth* criteria Eq. (8):

$$\underline{x_k} = \frac{1}{n} \sum_{i=1}^{m} x_{ik}; \quad k = 1, \dots, n. \quad (8)$$

The Pearson correlation coefficient captures linear correlations.

Step 4.2.4. The standard deviation of each criterion is estimated by Eq. (9):

$$\sigma_j = \sqrt{\frac{1}{n-1}\sum_{i=1}^{m} \left(x_{ij} - \underline{x_j}\right)^2}; \quad j = 1, \ldots, n \tag{9}$$

Step 4.2.5. The index of the jth criteria, Ej, is evaluated by Eq. (10)

$$E_j = \sigma_j \sum_{k=1}^{n} \left(1 - \rho_{jk}\right); \ j = 1, \ldots, n. \tag{10}$$

Step 4.2.6. The weights of the criteria are determined by Eq. (11)

$$W_{1Xj}^{C} = \frac{E_j}{\sum_{j=1}^{n} E_j}; \ j = 1, \ldots, n.$$
(11)

Finally, the ranking of the weights of the criteria is obtained. The ranking identifies the importance given to each criterion.

Step 5 – Definition of the lower and upper limits of the weights per criterion

After generating the weights of each criterion using the Entropy and CRITIC methods, which constitute the objective methods, the model opens the door to input weights from subjective methods, which can be obtained by a single decision maker or a group of decision-makers, with or without the use of subjective methods (Ayan, Abacıoğlu and Basilio, 2023) such as AHP; SAPEVO-M; FUCOM; MEREC among others.

In this step, we define the lower-limit vector. $Ll_{1Xj}$ where criterion j will store the smallest weight value obtained from the set of values formed by $\{W_j^E, W_j^C, W_j^{DM}\}$, as shown in Eq.(12)

$$Ll_{1Xj} = Min\{W_j^E, W_j^C, W_j^{DM}\}; \ j = 1, \ldots, n. \tag{12}$$

To illustrate this process, we have constructed the Table 2 below, with fictitious data, to illustrate Equation (12). The green colour represents the minimum points for each set of weights.

*Table 2 Lower limit definition statement*

| W \ C | $C_1$ | $C_2$ | $C_3$ |
|---|---|---|---|
| $W_j^E$ | 0,34 | 0,45 | 0,21 |
| $W_j^C$ | 0,23 | 0,40 | 0,37 |
| $W_j^{DM}$ | 0,41 | 0,37 | 0,22 |
| $Ll_{1Xj}$ | 0,23 | 0,37 | 0,21 |

we will define the upper limit vector. $Ul_{1Xj}$, which for each criterion j will store the highest weight value obtained from the set of values formed by $\{W_j^E, W_j^C, W_j^{DM}\}$, as shown in Eq.(13)

$$Ul_{1Xj} = Max\{W_j^E, W_j^C, W_j^{DM}\}; \ j = 1, \ldots, n. \tag{13}$$

To exemplify this process, we have constructed Table 3 below with fictitious data to illustrate Equation (13). In this case, the blue colour highlights the maximum point of each weight set.

*Table 3 Upper limit definition statement*

| W \ C | $C_1$ | $C_2$ | $C_3$ |
|---|---|---|---|
| $W_j^E$ | 0,34 | 0,45 | 0,21 |
| $W_j^C$ | 0,23 | 0,40 | 0,37 |
| $W_j^{DM}$ | 0,41 | 0,37 | 0,22 |
| $Ul_{1Xj}$ | 0,41 | 0,45 | 0,37 |

It should be made clear to the reader that the variable $W_j^{DM}$ corresponds to the insertion of the weight from outside the model, as shown in Figure 1. This insertion can come from subjective weights from the decision-makers, or group of decision-makers, or just the input of other objective weight generation methods, such as the AHP. The following Table 4 summarises the process of obtaining the upper and lower limits for each criterion.

*Table 4 Statement of upper and lower limits.*

| Limit \ Criteria | $C_1$ | $C_2$ | $C_3$ |
|---|---|---|---|
| $Ul_{1Xj}$ | 0,41 | 0,45 | 0,37 |
| $Ll_{1Xj}$ | 0,23 | 0,37 | 0,21 |

The example illustrates that the choices of upper and lower limits for each criterion can be made up of values produced by each of the methods. The predominance of any of the methods used cannot be guaranteed.

**Step 6 – Random generation of "t" sets of weights by criteria**

In this phase, the Randomised Weight Matrix RWm of dimension t x n will be generated. Where "t" is a value entered by the decision-maker in the problem, which corresponds to the number of sets of weights to be generated in the model, and "n" is the number of criteria in the model. This stage is one of the model's innovations. In the cases found in the literature, when more than one method is used to obtain the weights of the criteria, mathematical operations are usually carried out to obtain a single set of weights for the criteria. In this model, RWm will allow you to obtain "t" sets of weights, which, when applied to the TOPSIS method, will result in t final rankings. In reality, this process corresponds to a sensitivity analysis included in the model. Where "t" is the total number of rows, corresponding to the total number of iterations inserted in the model by the decision maker. Where n is the total number of columns of the matrix. The RWm matrix is obtained by generating different random numbers limited for each criterion by the limits. $\boldsymbol{Ll_j}$ and $\boldsymbol{Ul_j}$ , as shown in Eq. (14):

$$\boldsymbol{RWm_{ij}} = \left(\left(\boldsymbol{Ul_j} - \boldsymbol{Ll_j}\right) * \boldsymbol{Rnd}\right) + \boldsymbol{Ll_j}); \ \forall\, \boldsymbol{i = 1 \dots t}; \boldsymbol{And\ j = 1 \dots n}. \quad (14)$$

Next, the matrix $\boldsymbol{RWm_{ij}}$ is normalized by Eq. (15):

$$\boldsymbol{RWm^n_{ij}} = \frac{x_{ij}}{\sum_{j=1}^{n} x_{ij}}; \ \forall\, \boldsymbol{i = 1 \dots t}; \boldsymbol{And\ j = 1 \dots n}. \quad (15)$$

Step 7 – Generation of "t" Ranking with the TOPSIS method

Hwang and Yoon (Hwang and Yoon, 1981) developed in 1981 a Technique for Order Performance by Similarity to Ideal Solution (TOPSIS). According to this technique, the best alternative should be the one that is as close to the positive ideal solution as possible and the most distant from the negative ideal solution. The positive ideal solution is the one that maximizes the benefit criteria and minimizes the cost criteria; the negative ideal solution maximizes the cost criteria and minimizes the benefit criteria. In short, the positive ideal solution is made up of all the best achievable values for the benefit criteria, since the negative ideal solution consists of all worst achievable cost criteria.

Decision D Matrix, composed of *m* alternatives evaluated by *n* criteria (or attributes), is described by:

$$D = \begin{pmatrix} X_{11} \dots X_{12} \dots X_{13} \dots\dots\dots\dots X_{1n} \\ X_{21} \dots X_{22} \dots X_{23} \dots\dots\dots\dots X_{2n} \\ X_{31} \dots X_{32} \dots X_{33} \dots\dots\dots\dots X_{3n} \\ \dots\dots\dots\dots\dots\dots\dots\dots\dots\dots \\ \dots\dots\dots\dots X_{ij} \dots\dots\dots\dots \\ \dots\dots\dots\dots\dots\dots\dots\dots\dots\dots \\ X_{m1} \dots X_{m2} \dots X_{m3} \dots\dots\dots\dots X_{mn} \end{pmatrix}$$

$A_1, A_2, \dots, A_m$ are viable alternatives; $C_1, C_2, \dots, C_n$ are criteria; $X_{ij}$ indicates the performance of the alternative $A_{ij}$ according to $C_j$.

Step 7.1: Normalize the matrix in order to transform it into a dimensionless matrix so that it is possible to make a comparison between the several criteria. Matrix D is normalized for each criterion $C_j$

Through $p_{ij} = \frac{X_{ij}}{MAX\ \overline{x_i}}$ , whereas j= 1,....n and $\overline{x_i}$ represents the maximum value for $x_i$ for each criterion $C_j$.

Step 7.2: The calculation of weights for each one of the criteria. The vector of W weight composed of individual weights $W_j$ (j = 1,...,n) for each $C_j$ satisfies $\sum_{j=1}^{n} W_j = 1$ . Each criterion j will have a given weight.

$$W = \lfloor W_1, W_2, \dots\dots\dots, W_j, \dots\dots\dots W_n \rfloor$$

$$D_n = \begin{pmatrix} p_{11} \cdots p_{12} \cdots p_{13} \cdots\cdots\cdots p_{1n} \\ p_{21} \cdots p_{22} \cdots p_{23} \cdots\cdots\cdots p_{2n} \\ p_{31} \cdots p_{32} \cdots p_{33} \cdots\cdots\cdots p_{3n} \\ \cdots\cdots\cdots\cdots\cdots\cdots\cdots\cdots \\ \cdots\cdots\cdots p_{ij} \cdots\cdots\cdots \\ \cdots\cdots\cdots\cdots\cdots\cdots\cdots\cdots \\ p_{m1} \cdots p_{m2} \cdots p_{m3} \cdots\cdots\cdots p_{mn} \end{pmatrix}$$

Step 7.3 – Calculation of entropy (amount of information about the decision matrix)

$$e_j = -\frac{1}{\ln(m)} \sum_{i=1}^{m} P_{ij} \ln\left(p_{ij}\right) \tag{16}$$

And "m" stands for the number of alternatives, i=1...n, j=1...m

Step 7.4: The degree of diversity of information inside each one criterion is calculated according to

$$dj = 1 - ej \tag{17}$$

Step 7.5: The weight for each criterion by entropy method is calculated by,

$$W_j = \frac{d_j}{\sum_{j=1}^{n} d_j} \tag{18}$$

Step 7.6: Calculation is needed for the normalized values according to the weight.

$$V_{ij} = W_j * p_{ij} \tag{19}$$

This way, a new matrix of normalized decision also obtained from the weight of each criterion $D_{np}$ represents the relative performance of alternatives and might be described by:

$$D_{np} = \begin{pmatrix} V_{11} \cdots V_{12} \cdots V_{13} \cdots\cdots\cdots V_{1n} \\ V_{21} \cdots V_{22} \cdots V_{23} \cdots\cdots\cdots V_{2n} \\ V_{31} \cdots V_{32} \cdots V_{33} \cdots\cdots\cdots V_{3n} \\ \cdots\cdots\cdots\cdots\cdots\cdots\cdots\cdots \\ \cdots\cdots\cdots V_{ij} \cdots\cdots\cdots \\ \cdots\cdots\cdots\cdots\cdots\cdots\cdots\cdots \\ V_{m1} \cdots V_{m2} \cdots V_{m3} \cdots\cdots\cdots V_{mn} \end{pmatrix}$$

Step 7.7: Identify the positive ideal solutions $A^+$ (benefits) and the negative ideal solutions (costs) $A^-$ following these structures:

$$A^+ = \left\{V_1^*, \ldots\ldots, V_j^*, \ldots\ldots. V_n^*\right\} = \left\{\left(\max_j V_{ij} \mid \mathrm{j} = 1, \ldots, \mathrm{n}\right) \mid \mathrm{i} = 1, \ldots, \mathrm{m}\right\}$$

$$A^- = \left\{V_1^-, \ldots\ldots, V_j^-, \ldots\ldots. V_n^-\right\} = \left\{\left(\min_j V_{ij} \mid \mathrm{j} = 1, \ldots, \mathrm{n}\right) \mid \mathrm{i} = 1, \ldots, \mathrm{m}\right\}$$

Step 7.8: Calculation of Euclidean distances between $A_i$ e $A^+$ (benefits) and $A_i$ e $A^-$ (costs) adopting the following structures:

$$d^+ = \sqrt{\sum_{j=1}^{n} W_j \left(V_{ij} - V_j^*\right)^2} \quad .............. \quad .(20)$$

$$d^- = \sqrt{\sum_{j=1}^{n} W_j \left(V_{ij} - V_j^-\right)^2} \quad (21)$$

Step 7.9: Calculation of a relative similarity $\xi_i$ is required for each alternative $A_i$ in relation to the positive ideal solution $A^+$ according to:

$$\xi_i = \frac{d_i^-}{d_i^- + d_i^+} \quad (22)$$

Step 7.10: Ranking according to the relative similarity. The best alternatives are those that have the highest values of $\boldsymbol{\xi_i}$ and must be chosen since they are closer to the positive ideal solution.

Step 8 – Definition of final ranking

In this step, we present the second novelty of this new method. In step 6, we present the matrix $\boldsymbol{RWm_{ij}^n}$. The matrix $\boldsymbol{RWm_{ij}^n}$ contains t sets of weights per criterion. The innovation point of this method is to generate t sets of rankings as different sets of weights are used, varying within the range of weights for each criterion, as dealt with in Step 5. In this sense, $\boldsymbol{\xi_i}$ is transformed into an ordinal value. The $\boldsymbol{\xi_i}$ is sorted in descending order, being assigned 1st place to alternative ($\boldsymbol{A_i}$) that has the highest $\boldsymbol{\xi_i}$, and so on until the last alternative m. The final ranking matrix FRm is of dimension m x t, where m is the number of rows composed of each alternative ($\boldsymbol{A_i}$). Where t is the number of columns representing the ranking generated by the TOPSIS method for each iteration. $\boldsymbol{a_{ij}}$ is the ordinal value of the ranking that alternative j obtained in iteration i. As shown in Eq. (23):

$$\boldsymbol{FRm_{mxt}} = \begin{bmatrix} \boldsymbol{r_{11}} & \cdots & \boldsymbol{r_{1t}} \\ \boldsymbol{r_{21}} & \cdots & \boldsymbol{r_{2t}} \\ \cdots & \cdots & \cdots \\ \boldsymbol{r_{51}} & \cdots & \boldsymbol{r_{5t}} \\ \cdots & \cdots & \cdots \\ \boldsymbol{r_{m1}} & \cdots & \boldsymbol{r_{mt}} \end{bmatrix}$$

$$\boldsymbol{FRm_{ij}} = \sum_{i=1}^{m} \sum_{j=1}^{t} \boldsymbol{\xi_i}, \forall\, \boldsymbol{i = 1, 2, \ldots, m \;\; and\, j = 1, 2, \ldots, t.} \quad (23)$$

Then, the value of each rank-ordering $\boldsymbol{a_{ij}}$ will be replaced by a score, as follows: 1st = m, 2nd=(m-1), ..., nth=(m-(m-1). Thus, the final ranking vector FRv of dimension "i", which corresponds to the total number of alternatives. The final position of each alternative will be obtained using the MODE statistical measure for the set of positions assigned in the execution of "t" iterations of each alternative. As shown in Eq.(24):

$$\boldsymbol{FRv_i} = \boldsymbol{MODE}\left(\sum_{j=1}^{t} \boldsymbol{FRm_{ij}}\right), \boldsymbol{i = 1, 2, \ldots, m, and\, j = 1, 2, \ldots, t.} \quad (24)$$

The final ranking will be obtained in descending order among the ordinal value of each alternative "i" of the vector $\boldsymbol{FRv_i}$.

## 4. Method validation

In this section, data from the research conducted by Ayan and Abacıoğlu (Ayan and Abacıoğlu, 2022) is utilized. The study introduced a multicriteria decision-making (MCDM) approach to assess companies' performance in relation to user-generated content (UGC) metrics, which were defined and calculated as ratios derived from each tweet's sentiment type (positive, negative, or neutral) and relevant metrics such as the number of tweets, retweets, favorites, and reach. In the current analysis, the EC-TOPSIS tool(Ayan and Abacıoğlu, 2022), developed in Python and available at https://pypi.org/project/ec-topsis/, is applied to evaluate the dataset.

*Data:*

*Criteria and Alternatives:*

Twitter, a microblogging service, was chosen as the current study's social media platform sample(Ayan and Abacıoğlu, 2022). Twitter plays an essential role for both institutions and individual users. It enables companies to reach out to customers and share information. Individuals or consumers can contact companies and generate content for them (Chu, Chen and Sung, 2016). For company selection, the Boom Social platform (Boomsocial, 2018) was used, which is a website that allows various comparisons and reports on social media. Various industries' ranking lists on this platform were examined. Among the shopping industry's companies, three industries, which included two competitors with the highest number of followers, were determined. In the current study, cosmetics, marketplace, and electronic industries were selected and six companies were evaluated. Cosmetics companies run retail stores where they sell cosmetics and personal care products. The marketplace industry consists of e-commerce companies that sell multi-category products exclusively online, whereas the electronic industry involves multi-channel retail chains where electronic or technology products are sold. To protect the anonymity of the selected companies, they are labeled C1, C2, M1, M2, E1, and E2 where "C" for cosmetics, "M" for marketplace, and "E" for electronics. For educational purposes, we'll rename the alternatives as follows: A1=C1; A2=C2; A3=M1; A4=M2; A5=E1; and A6=E2.

Retweets and reach (the number of followers of users who retweet the company's tweets) were decided by researchers Ayan and Abacıoğlu (Ayan and Abacıoğlu, 2022) as criteria based on Malhotra (Sterne, 2010). The cost or benefit of the criteria was defined according to the types of sentiment. Negative sentiment indicates the cost of the criteria, while positive sentiment shows the benefit of the criteria. Neutral sentiment is also considered to be the benefit of the criteria, such as sharing, retweeting, and liking tweets that do not involve negative statements is beneficial for companies' awareness, engagement, and word of mouth (Hoffman and Fodor, 2020). In addition, being commented on social media can be seen as a strength for companies (Capatina *et al.*, 2018). C1, C2, and C3 are calculated by dividing the number of positive, negative, and neutral tweets by the total number of tweets. The other criteria, on the other hand, deal with the same rate account by considering the number of retweets (C4, C5, and C6), favorites (C7, C8, and C9), and reach (C10, C11, and C12) instead of the number of tweets.

Table 5 outlines the 6 alternatives, representing the selected companies, and the 12 criteria derived from the original model introduced by Ayan and Abacıoğlu (Ayan and Abacıoğlu, 2022). These criteria include various Twitter-based metrics, as described above. The type (cost-min or benefit-max) of each criterion is also indicated.

*The Decision Matrix*

The decision matrix, which constitutes the initial input for the EC-TOPSIS tool, is presented in Table 6.

*Table 5 Alternatives and criteria established for analysis*

| Alternatives | Criteria | | Cost or Benefit |
|---|---|---|---|
| A1 | C1 | Positive tweets ratio | Benefit (max) |
| A2 | C2 | Negative tweets ratio | Cost (min) |
| A3 | C3 | Neutral tweets ratio | Benefit (max) |
| A4 | C4 | Retweets ratio of positive tweets | Benefit (max) |
| A5 | C5 | Retweets ratio of negative tweets | Cost (min) |
| A6 | C6 | Retweets ratio of neutral tweets | Benefit (max) |
| | C7 | Favorites ratio of positive tweets | Benefit (max) |
| | C8 | Favorites ratio of negative tweets | Cost (min) |
| | C9 | Favorites ratio of neutral tweets | Benefit (max) |
| | C10 | Reach ratio of positive tweets | Benefit (max) |
| | C11 | Reach ratio of negative tweets | Cost (min) |
| | C12 | Reach ratio of neutral tweets | Benefit (max) |

Note: Data obtained from (Ayan and Abacıoğlu, 2022)

Table 6 **Decision matrix of evaluation of the companies' social media metrics**

| | Criteria | | | | | | | | | | | |
|---|---|---|---|---|---|---|---|---|---|---|---|---|
| Alternatives | C1 | C2 | C3 | C4 | C5 | C6 | C7 | C8 | C9 | C10 | C11 | C12 |
| A1 | 0.315 | 0.141 | 0.544 | 0.323 | 0.047 | 0.630 | 0.219 | 0.060 | 0.722 | 0.198 | 0.063 | 0.739 |
| A2 | 0.299 | 0.132 | 0.569 | 0.270 | 0.132 | 0.598 | 0.061 | 0.040 | 0.899 | 0.154 | 0.067 | 0.779 |
| A3 | 0.044 | 0.323 | 0.633 | 0.006 | 0.206 | 0.788 | 0.037 | 0.058 | 0.906 | 0.004 | 0.022 | 0.974 |
| A4 | 0.056 | 0.069 | 0.875 | 0.000 | 0.009 | 0.991 | 0.003 | 0.005 | 0.992 | 0.009 | 0.005 | 0.986 |
| A5 | 0.013 | 0.086 | 0.901 | 0.001 | 0.019 | 0.979 | 0.013 | 0.021 | 0.966 | 0.001 | 0.001 | 0.998 |
| A6 | 0.039 | 0.346 | 0.615 | 0.056 | 0.268 | 0.677 | 0.001 | 0.004 | 0.995 | 0.002 | 0.026 | 0.972 |

**Note**: The Twitter metrics based on UGC were calculated by considering tweets generated and shared about companies, the number of retweets that each tweet received, the number of favorites that each tweet received, and the number of followers that each tweet's owner (user) had. Data obtained from (Ayan and Abacıoğlu, 2022)

## 5. Results and Discussion

In this section, we will present the results obtained with the EC-TOPIS method and compare them with the original results obtained with the CRITIC-ARAS method and CRITIC-COPRAS. We will then use other objective methods for obtaining criteria weights, apply them to other ranking methods, and compare the results with those of EC-TOPSIS to test the robustness of the method proposed in this paper.

### 5.1 EC-TOPSIS Results

Stages 1, 2, and 3 of the EC-TOPSIS model were obtained using data from the research by Ayan and Abacıoğlu (Ayan and Abacıoğlu, 2022). The proposed model enables the decision maker to utilize weights objectively generated by the Entropy and CRITIC methods within the EC-TOPSIS framework. Additionally, the decision maker has the option to incorporate an external set of weights, which will serve as parameters for determining the upper and lower limits of the weight ranges generated by these methods. In this case, a uniform weighting strategy was implemented, allocating an equal 0.05 weight to each of the 12 criteria. This approach ensures a balanced evaluation by preventing any single criterion from dominating the decision-making process, while still maintaining the potential for alternative weighting methods such as expert consultation or Analytic Hierarchy Process (AHP).

The code below defines the parameters required for implementing the method:

```
criterion_type = ['max', 'min', 'max', 'max', 'min', 'max', 'max', 'min', 'max', 'max', 'min', 'max']

iterations      = 10,000

# OPTIONAL: User-defined Custom Weigths

custom_sets = [

    [0.05] * 12

]

# Dataset

dataset = np.array([

    [0.315, 0.141, 0.544, 0.323, 0.047, 0.630, 0.219, 0.060, 0.722, 0.198, 0.063, 0.739],

    [0.299, 0.132, 0.569, 0.270, 0.132, 0.598, 0.061, 0.040, 0.899, 0.154, 0.067, 0.779],

    [0.044, 0.323, 0.633, 0.006, 0.206, 0.788, 0.037, 0.058, 0.906, 0.004, 0.022, 0.974],

    [0.056, 0.069, 0.875, 0.000, 0.009, 0.991, 0.003, 0.005, 0.992, 0.009, 0.005, 0.986],

    [0.013, 0.086, 0.901, 0.001, 0.019, 0.979, 0.013, 0.021, 0.966, 0.001, 0.001, 0.998],

    [0.039, 0.346, 0.615, 0.056, 0.268, 0.677, 0.001, 0.004, 0.995, 0.002, 0.026, 0.972]

])
```

After entering the criteria, alternatives, and evaluation matrix of the model into the EC-TOPSIS tool, the following steps were initiated. Table 7 presents the weights generated internally by the EC-TOPSIS method, alongside the incorporated external weights. The outcome of these steps is determining the upper and lower limits of the weight ranges for each criterion, which will be used to calculate the total number of interactions required to establish the model's final ranking.

```
# Run EC-TOPSIS

ect = ec_topsis(

                dataset = dataset,

                criterion_type = criterion_type,

                custom_sets = custom_sets,

                iterations= iterations

               )

# Show Weights Lower and Upper Bounds

df = ect.weights_df

data_table.DataTable(df.round(3), num_rows_per_page = 15)
```

*Table 7 Output with the weights generated and the definition of the upper and lower limits of the model's weight ranges.*

| Weight Name | g1 | g2 | g3 | g4 | g5 | g6 | g7 | g8 | g9 | g10 | g11 | g12 |
|---|---|---|---|---|---|---|---|---|---|---|---|---|
| Entropy | 0.092 | 0.029 | 0.004 | 0.155 | 0.112 | 0.004 | 0.148 | 0.096 | 0.001 | 0.173 | 0.185 | 0.001 |
| Critic | 0.101 | 0.064 | 0.067 | 0.104 | 0.061 | 0.069 | 0.097 | 0.078 | 0.085 | 0.104 | 0.076 | 0.094 |
| Custom Weights 1 | 0.083 | 0.083 | 0.083 | 0.083 | 0.083 | 0.083 | 0.083 | 0.083 | 0.083 | 0.083 | 0.083 | 0.083 |
| Lower | 0.083 | 0.029 | 0.004 | 0.083 | 0.061 | 0.004 | 0.083 | 0.078 | 0.001 | 0.083 | 0.076 | 0.001 |
| Upper | 0.101 | 0.083 | 0.083 | 0.155 | 0.112 | 0.083 | 0.148 | 0.096 | 0.085 | 0.173 | 0.185 | 0.094 |

*Note:* Data output generated by the EC-TOPSIS tool in python.

After determining the weight range limits for each criterion, the EC-TOPSIS tool generates multiple weight sets to be applied to the ranking model. In this validation, we conducted 10,000 iterations to explore diverse weight configurations. Table 8 presents the comprehensive results obtained from these systematically generated weight sets, providing a robust analysis of the ranking model's sensitivity to different weighting scenarios.

```
# Weights Matrix

dw = ect.df_w

data_table.DataTable(dw.round(3), num_rows_per_page = 15)
```

*Table 8 Table with n iterations of the generated criteria bands.*

| Index | g1 | g2 | g3 | g4 | g5 | g6 | g7 | g8 | g9 | g10 | g11 | g12 |
|---|---|---|---|---|---|---|---|---|---|---|---|---|
| Iteration 1 | 0.097 | 0.042 | 0.061 | 0.103 | 0.076 | 0.035 | 0.138 | 0.094 | 0.037 | 0.137 | 0.095 | 0.002 |
| Iteration 2 | 0.091 | 0.068 | 0.067 | 0.091 | 0.08 | 0.013 | 0.144 | 0.092 | 0.03 | 0.105 | 0.124 | 0.054 |
| Iteration 3 | 0.084 | 0.077 | 0.02 | 0.153 | 0.067 | 0.082 | 0.13 | 0.092 | 0.047 | 0.143 | 0.125 | 0.05 |
| Iteration 4 | 0.087 | 0.078 | 0.076 | 0.125 | 0.092 | 0.052 | 0.087 | 0.082 | 0.025 | 0.118 | 0.166 | 0.007 |
| Iteration 5 | 0.096 | 0.076 | 0.032 | 0.145 | 0.093 | 0.02 | 0.116 | 0.095 | 0.081 | 0.126 | 0.175 | 0.084 |
| Iteration 6 | 0.095 | 0.039 | 0.021 | 0.149 | 0.072 | 0.03 | 0.09 | 0.09 | 0.055 | 0.163 | 0.113 | 0.061 |
| Iteration 7 | 0.097 | 0.029 | 0.065 | 0.098 | 0.065 | 0.034 | 0.13 | 0.084 | 0.01 | 0.096 | 0.151 | 0.023 |
| Iteration 8 | 0.099 | 0.033 | 0.011 | 0.119 | 0.089 | 0.051 | 0.119 | 0.093 | 0.053 | 0.133 | 0.118 | 0.053 |
| Iteration 9 | 0.096 | 0.076 | 0.013 | 0.151 | 0.075 | 0.027 | 0.109 | 0.089 | 0.079 | 0.092 | 0.16 | 0.018 |
| Iteration 10 | 0.097 | 0.058 | 0.051 | 0.154 | 0.1 | 0.042 | 0.087 | 0.087 | 0.075 | 0.129 | 0.155 | 0.061 |
| Iteration 11 | 0.086 | 0.066 | 0.043 | 0.147 | 0.101 | 0.074 | 0.103 | 0.084 | 0.034 | 0.118 | 0.165 | 0.059 |
| Iteration 12 | 0.097 | 0.061 | 0.073 | 0.138 | 0.103 | 0.079 | 0.084 | 0.091 | 0.013 | 0.145 | 0.167 | 0.088 |
| Iteration 13 | 0.1 | 0.075 | 0.073 | 0.131 | 0.078 | 0.076 | 0.134 | 0.094 | 0.07 | 0.126 | 0.089 | 0.041 |
| Iteration 14 | 0.095 | 0.066 | 0.006 | 0.129 | 0.073 | 0.043 | 0.135 | 0.078 | 0.079 | 0.155 | 0.121 | 0.043 |
| Iteration 15 | 0.084 | 0.034 | 0.059 | 0.119 | 0.077 | 0.066 | 0.124 | 0.082 | 0.078 | 0.105 | 0.094 | 0.085 |
| ... | ... | ... | ... | ... | ... | ... | ... | ... | ... | ... | ... | ... |
| Iteration 9991 | 0.099 | 0.079 | 0.081 | 0.112 | 0.078 | 0.012 | 0.124 | 0.084 | 0.035 | 0.088 | 0.109 | 0.019 |
| Iteration 9992 | 0.098 | 0.03 | 0.083 | 0.143 | 0.071 | 0.083 | 0.143 | 0.08 | 0.05 | 0.128 | 0.123 | 0.014 |
| Iteration 9993 | 0.09 | 0.07 | 0.049 | 0.097 | 0.108 | 0.058 | 0.131 | 0.081 | 0.055 | 0.147 | 0.104 | 0.044 |
| Iteration 9994 | 0.085 | 0.031 | 0.028 | 0.145 | 0.068 | 0.077 | 0.084 | 0.09 | 0.011 | 0.16 | 0.098 | 0.054 |
| Iteration 9995 | 0.09 | 0.057 | 0.067 | 0.084 | 0.075 | 0.026 | 0.145 | 0.095 | 0.007 | 0.086 | 0.095 | 0.069 |
| Iteration 9996 | 0.088 | 0.066 | 0.057 | 0.132 | 0.084 | 0.059 | 0.083 | 0.086 | 0.008 | 0.098 | 0.18 | 0.035 |
| Iteration 9997 | 0.096 | 0.078 | 0.078 | 0.126 | 0.077 | 0.038 | 0.11 | 0.083 | 0.047 | 0.103 | 0.185 | 0.021 |
| Iteration 9998 | 0.085 | 0.057 | 0.073 | 0.139 | 0.067 | 0.07 | 0.133 | 0.089 | 0.01 | 0.156 | 0.172 | 0.032 |
| Iteration 9999 | 0.098 | 0.063 | 0.01 | 0.115 | 0.079 | 0.017 | 0.121 | 0.09 | 0.025 | 0.138 | 0.157 | 0.056 |
| Iteration 10,000 | 0.099 | 0.05 | 0.04 | 0.09 | 0.11 | 0.056 | 0.146 | 0.088 | 0.056 | 0.119 | 0.097 | 0.082 |

*Note:* Data output generated by the EC-TOPSIS tool in python.

Beyond the tabular representation of iterations generated by the Python tool, the decision maker gains access to a graphical visualization that elucidates the weight distribution and range for each criterion. Figure 2 showcases an intermediate analytical output, offering a comprehensive visual interface for deeper insights into the weight allocation and distribution patterns, thereby enhancing the decision-making process through enhanced data interpretation.

```
# Weights Box Plot

ect.wm_boxplot(size_x = 15, size_y = 10)
```

Table 9 illustrates the n = 10,000 rankings generated in the model based on the weight sets presented in Table 8.

```
# Ranks Matrix

dr = ect.df_r

data_table.DataTable(dr.round(3), num_rows_per_page = 15)
```

*Table 9 Record of n = 10,000 ranks generated by EC-TOPSIS.*

| c | a1 | a2 | a3 | a4 | a5 | a6 |
|---|---|---|---|---|---|---|
| Iteration 1 | 1 | 2 | 6 | 3 | 4 | 5 |
| Iteration 2 | 1 | 2 | 6 | 3 | 4 | 5 |
| Iteration 3 | 1 | 2 | 6 | 3 | 4 | 5 |
| Iteration 4 | 1 | 2 | 6 | 3 | 4 | 5 |
| Iteration 5 | 1 | 2 | 6 | 3 | 4 | 5 |
| Iteration 6 | 1 | 2 | 6 | 3 | 4 | 5 |
| Iteration 7 | 1 | 2 | 6 | 3 | 4 | 5 |
| Iteration 8 | 1 | 2 | 6 | 3 | 4 | 5 |
| Iteration 9 | 1 | 2 | 6 | 3 | 4 | 5 |
| Iteration 10 | 1 | 2 | 6 | 3 | 4 | 5 |
| Iteration 11 | 1 | 2 | 6 | 3 | 4 | 5 |
| Iteration 12 | 1 | 2 | 6 | 3 | 4 | 5 |
| Iteration 13 | 1 | 2 | 6 | 3 | 4 | 5 |
| Iteration 14 | 1 | 2 | 6 | 3 | 4 | 5 |
| Iteration 15 | 1 | 2 | 6 | 3 | 4 | 5 |
| ... | ... | ... | ... | ... | ... | ... |
| Iteration 9991 | 1 | 2 | 6 | 3 | 4 | 5 |
| Iteration 9992 | 1 | 2 | 6 | 3 | 4 | 5 |
| Iteration 9993 | 1 | 2 | 6 | 3 | 4 | 5 |
| Iteration 9994 | 1 | 2 | 6 | 3 | 4 | 5 |
| Iteration 9995 | 1 | 2 | 6 | 3 | 4 | 5 |
| Iteration 9996 | 1 | 4 | 6 | 2 | 3 | 5 |
| Iteration 9997 | 1 | 3 | 6 | 2 | 4 | 5 |
| Iteration 9998 | 1 | 2 | 6 | 3 | 4 | 5 |
| Iteration 9999 | 1 | 2 | 6 | 3 | 4 | 5 |
| Iteration 10,000 | 1 | 2 | 6 | 3 | 4 | 5 |

Note: Data output generated by the EC-TOPSIS tool in python.

The developed Python tool offers a sophisticated visual analytics solution, enabling researchers and decision-makers to objectively examine ranking variations across solution alternatives through comprehensive iterative analysis. Figure 3 illustrates the graphical interpretation of ranks documented in Table 9, providing a clear visualization of how weight fluctuations systematically impact alternative rankings. This innovative approach effectively implements an embedded sensitivity analysis mechanism within the EC-TOPSIS methodological framework.

```
# Plot Ranks

ect.plot_rank_freq(size_x = 15, size_y = 10)
```

The EC-TOPSIS tool culminates in a comprehensive final ranking process, generating two distinct analytical outputs. The first provides a structured list of alternatives with their definitive ranking positions, while the second offers a sophisticated graphical solution. This visual representation strategically orders alternatives based on their final rank and incorporates a boxplot analysis, enabling a nuanced exploration of ranking distribution dynamics, as illustrated in Figures 4 and 5. This dual-output approach enhances decision-making transparency and provides a robust understanding of the ranking methodology's performance.

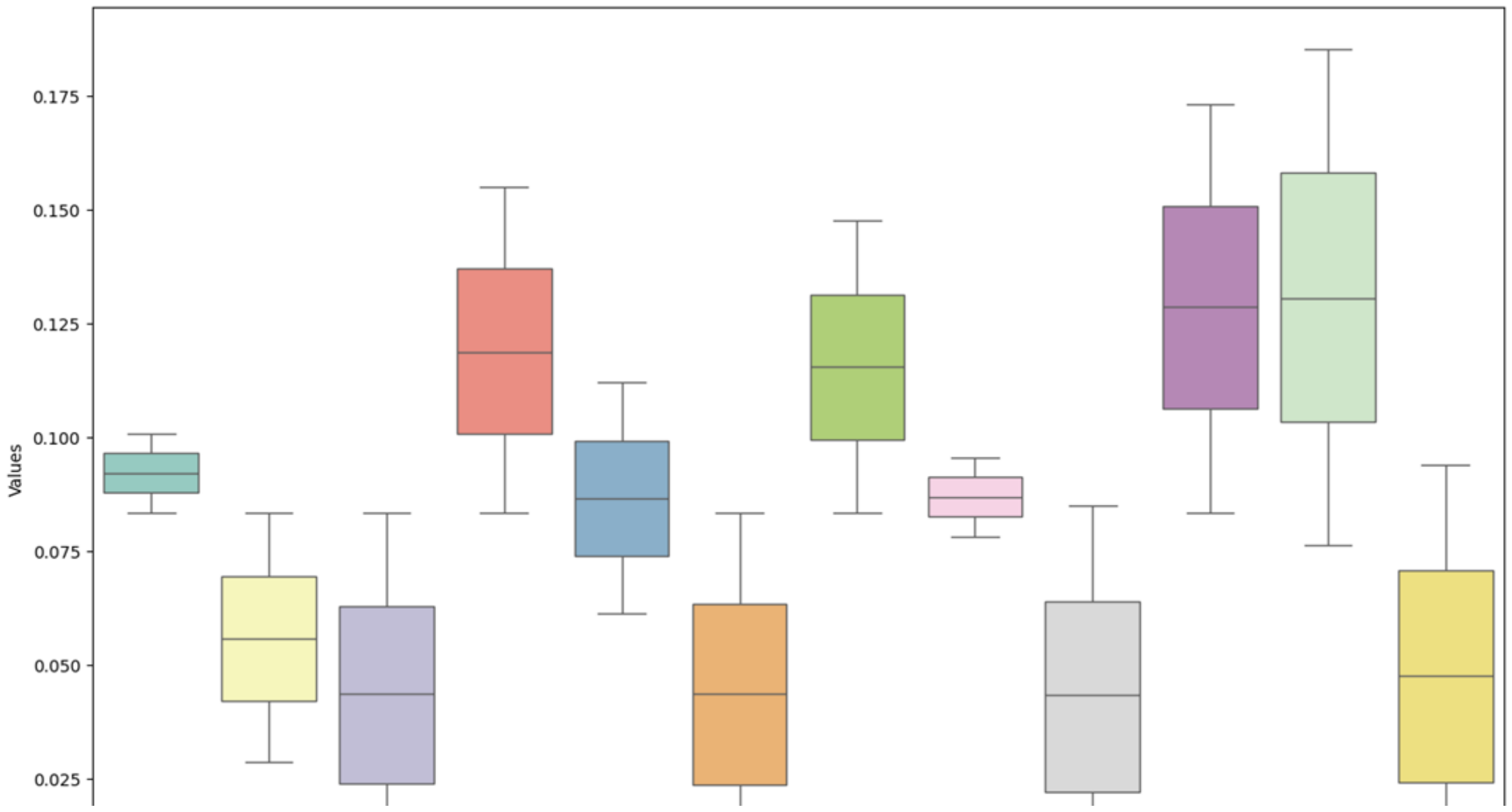

*Figure 2 Graphical representation of the variation in the weight range for each criterion. Note: graphical output is generated by the EC-TOPSIS tool in Python.*

```
# TOPSIS Box Plot

ect.topsis_boxplot(size_x = 15, size_y = 10)

# Plot TOPSIS Rank
```

*ect.ranking()*

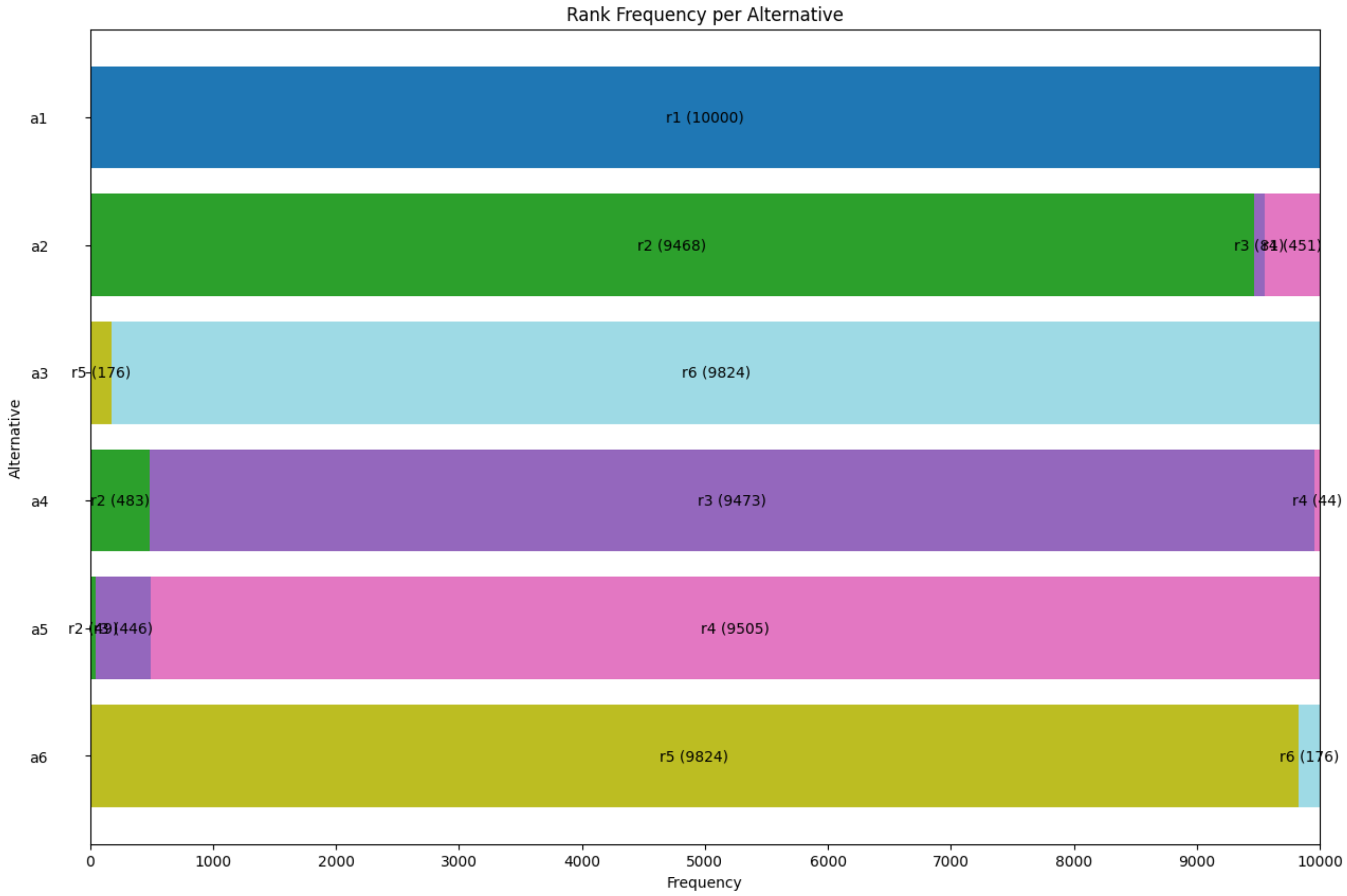

*Figure 3 Graphical representation of the variation in the ranks occupied by the alternatives throughout the process.*

*Note:* graphical output is generated by the EC-TOPSIS tool in Python.

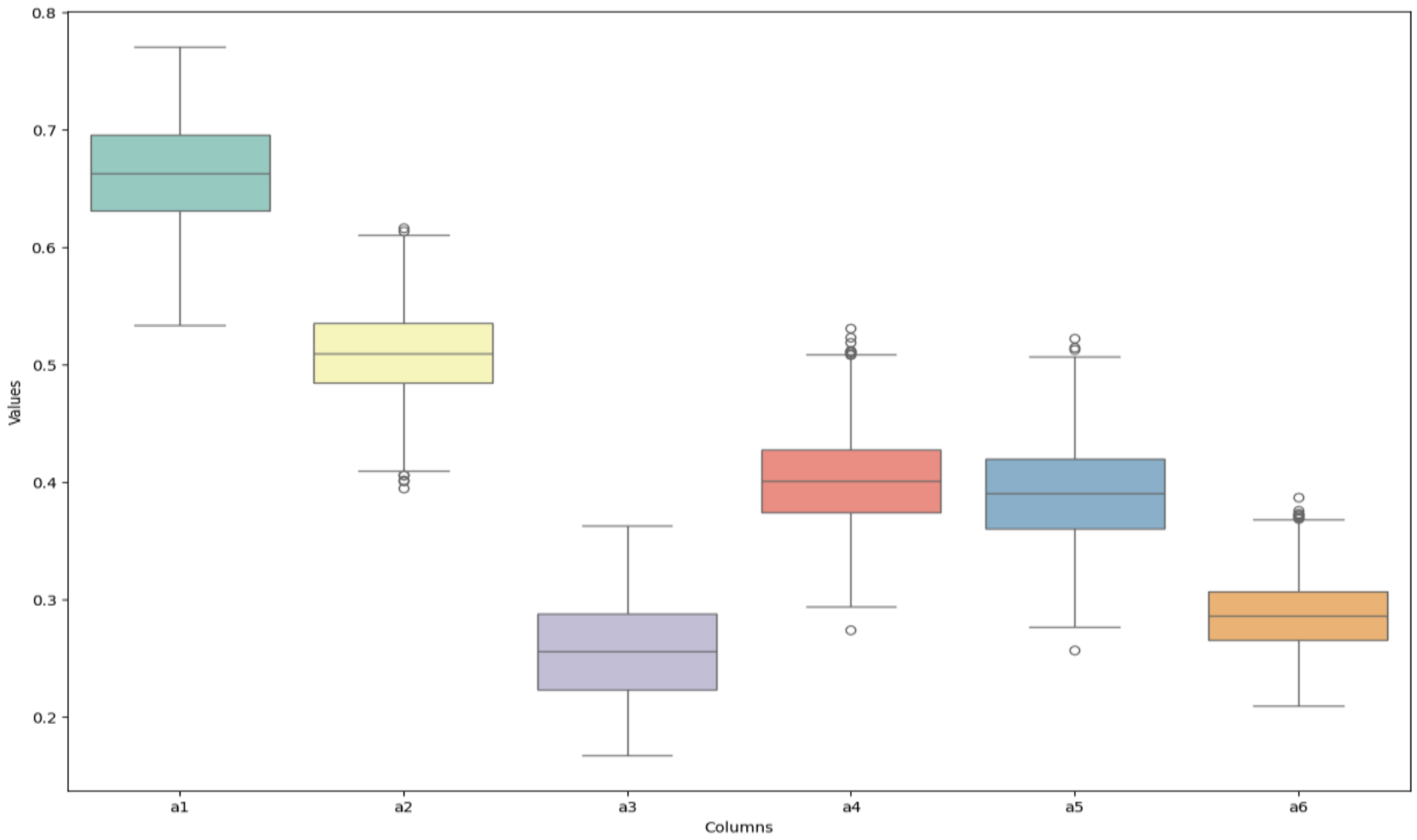

*Figure 4 Boxplot with the distribution of the final flow of EC-TOPSIS.*

*Note:* Graphical output is generated by the EC-TOPSIS tool in Python.

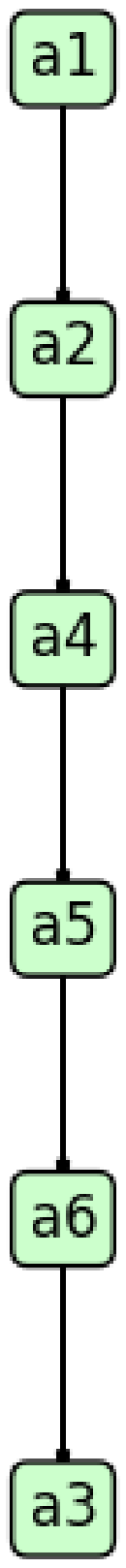

*Figure 5 Graphical representation of the final rank of the EC-TOPSIS method.*

*Note*: Graphical output is generated by the EC-TOPSIS tool in Python.

Additionally, the tool implements a rank mode functionality that systematically prints out each alternative's final ranking position. This feature allows for a clear, sequential enumeration of alternatives (a1, a2, a3, etc.) alongside their corresponding rank, providing an immediate and intuitive visualization of the ranking outcomes.

```
# Rank Mode

rank_mode = ect.sol_m

count = 1

for i in range(0, len(rank_mode)):

  print('a'+str(count)+':', rank_mode[i])

  count = count + 1
```

Output:

a1: [1]

a2: [2]

a3: [6]

a4: [3]

a5: [4]

a6: [5]

The method validation encompasses a comprehensive approach to substantiating the EC-TOPSIS results through an alternative methodological perspective. A supplementary validation strategy was implemented, deliberately employing a randomly selected weighting method and alternative ranking methods to facilitate a rigorous cross-comparative analysis. This methodological triangulation serves to enhance the robustness of the initial findings by introducing an orthogonal analytical framework that independently evaluates the decision matrix through distinctly different computational mechanisms.

By implementing alternative strategies, the research design intentionally introduces methodological variance. This approach allows for a critical examination of the EC-TOPSIS outcomes, providing a means to validate the consistency and reliability of the initial multi-criteria decision analysis through an independent computational lens. The comparative methodology serves not merely as a validation mechanism, but as a sophisticated cross-verification strategy that strengthens the overall methodological integrity of the research design.

In the original research paper (Ayan and Abacıoğlu, 2022), the analysis was conducted utilizing the CRITIC weighting method in conjunction with ARAS and COPRAS ranking techniques. Upon comparative analysis with EC-TOPSIS, the results demonstrated remarkable consistency. Specifically, CRITIC-COPRAS and EC-TOPSIS yielded identical outcomes, while CRITIC-ARAS produced nearly identical results, with only a different ranking that was consistently observed in the original study. To provide a more detailed analysis of the results, the EC-TOPSIS and CRITIC-COPRAS methods produced identical rankings (a1, a2, a4, a5, a6, and a3), while the CRITIC-ARAS method differed slightly by swapping the positions of the third and fourth alternatives, resulting in the ranking (a1, a2, a5, a4, a6, and a3). This highlights a high degree of consistency among the methods, with minor variations in specific rankings.

**5.2 IDOCRIW and MEREC-based Rankings (ARAS, COPRAS, EDAS, MARCOS, and TOPSIS)**

Given that EC-TOPSIS inherently incorporates the CRITIC method, the researchers sought to compare different weighting methods for further methodological triangulation. Consequently, considering the novel weighting methods, IDOCRIW and MEREC, were strategically selected as different weighting techniques to provide an additional layer of validation and methodological robustness. The ranking methods selected for the analysis were ARAS, COPRAS, EDAS, MARCOS, and TOPSIS. Since ARAS, COPRAS, and TOPSIS are already included in the existing comparisons, the novel ranking methods EDAS and MARCOS have also been selected for additional comparisons. All application codes can be accessed from the relevant GitHub repository: https://github.com/Valdecy/pyDecision?tab=readme-ov-file

The input parameter codes for all IDOCRIW and MEREC methods are as follows:

```
# Load Criterion Type: 'max' or 'min'

criterion_type = ['max', 'min', 'max', 'max', 'min', 'max', 'max', 'min', 'max', 'max', 'min', 'max']

# Dataset

dataset = np.array([

    [0.315, 0.141, 0.544, 0.323, 0.047, 0.630, 0.219, 0.060, 0.722, 0.198, 0.063, 0.739],

    [0.299, 0.132, 0.569, 0.270, 0.132, 0.598, 0.061, 0.040, 0.899, 0.154, 0.067, 0.779],

    [0.044, 0.323, 0.633, 0.006, 0.206, 0.788, 0.037, 0.058, 0.906, 0.004, 0.022, 0.974],
```

```
    [0.056, 0.069, 0.875, 0.000, 0.009, 0.991, 0.003, 0.005, 0.992, 0.009, 0.005, 0.986],
    [0.013, 0.086, 0.901, 0.001, 0.019, 0.979, 0.013, 0.021, 0.966, 0.001, 0.001, 0.998],
    [0.039, 0.346, 0.615, 0.056, 0.268, 0.677, 0.001, 0.004, 0.995, 0.002, 0.026, 0.972]
])
```

Subsequently, the IDOCRIW and MEREC functions were called to derive the criterion weights.

```
# Call IDOCRIW Function
weights = idocriw_method(dataset, criterion_type, verbose = False)
# Weigths
for i in range(0, weights.shape[0]):
  print('w(g'+str(i+1)+'): ', round(weights[i], 3))
```

IDOCRIW Output:

w(g1): 0.093

w(g2): 0.069

w(g3): 0.056

w(g4): 0.119

w(g5): 0.087

w(g6): 0.056

w(g7): 0.116

w(g8): 0.079

w(g9): 0.055

w(g10): 0.127

w(g11): 0.086

w(g12): 0.055

```
# Call MEREC Function
weights = merec_method(dataset, criterion_type)
# Weigths
for i in range(0, weights.shape[0]):
```

*print('w(g'+str(i+1)+'): ', round(weights[i], 3))*

MEREC Output:

w(g1): 0.053

w(g2): 0.028

w(g3): 0.008

w(g4): 0.57

w(g5): 0.053

w(g6): 0.009

w(g7): 0.088

w(g8): 0.043

w(g9): 0.008

w(g10): 0.077

w(g11): 0.056

w(g12): 0.007

Following the weight determination through IDOCRIW and MEREC methods, the derived weights were subsequently applied to five distinct ranking techniques: ARAS, COPRAS, EDAS, MARCOS, and TOPSIS. The input parameter codes for all the methods are as follows:

```
# Weights

weights = [the weight sets derived from IDOCRIW and MEREC methodologies were directly incorporated into the computational framework]

# Load Criterion Type: 'max' or 'min'

criterion_type = ['max', 'min', 'max', 'max', 'min', 'max', 'max', 'min', 'max', 'max', 'min', 'max']

dataset = np.array([

    [0.315, 0.141, 0.544, 0.323, 0.047, 0.630, 0.219, 0.060, 0.722, 0.198, 0.063, 0.739],

    [0.299, 0.132, 0.569, 0.270, 0.132, 0.598, 0.061, 0.040, 0.899, 0.154, 0.067, 0.779],

    [0.044, 0.323, 0.633, 0.006, 0.206, 0.788, 0.037, 0.058, 0.906, 0.004, 0.022, 0.974],

    [0.056, 0.069, 0.875, 0.000, 0.009, 0.991, 0.003, 0.005, 0.992, 0.009, 0.005, 0.986],

    [0.013, 0.086, 0.901, 0.001, 0.019, 0.979, 0.013, 0.021, 0.966, 0.001, 0.001, 0.998],

    [0.039, 0.346, 0.615, 0.056, 0.268, 0.677, 0.001, 0.004, 0.995, 0.002, 0.026, 0.972]

])
```

After calling the methods' functions (e.g., *# Call EDAS Function; rank = edas_method(dataset, criterion_type, weights, graph=True, verbose=True)*), the obtained ranking results are given in Fig 6.

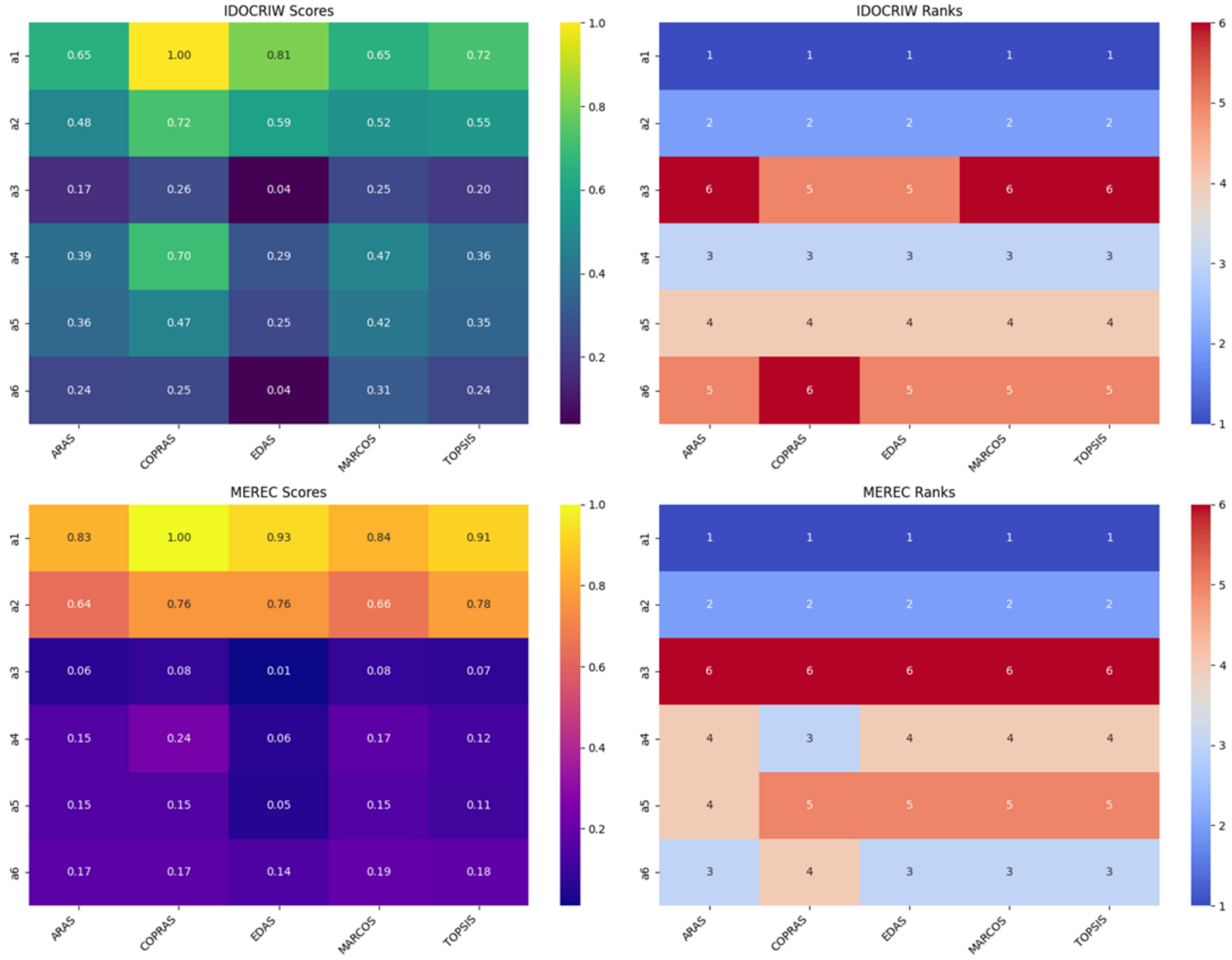

*Figure 6 . IDOCRIW and MEREC-based ranking comparisons for ARAS, COPRAS, EDAS, MARCOS, and TOPSIS, respectively.*

**Note: The graphical output is generated using Python code developed by the authors.**

In this section, we validate the proposed method using data from the social media problem, comparing the results of the EC-TOPSIS outputs with those of the original CRITIC-COPRAS and CRITC-ARAS methods, with only two changes in the ranking of the CRITIC-ARAS method, preserving the initial positions. Secondly, we used two other methods to obtain the weights of the criteria - MEREC and IDOCRIW - integrated with the ARAS, COPRAS, EDAS, MARCOS, and TOPSIS methods, the results of which are illustrated in Figure 6. We can see that there are few changes in the rankings, with the top positions being preserved. In general, we can infer that the EC-TOPSIS method can be used as a safe alternative by decision-makers to replace the compared methods. As perfection is a goal to be pursued, we cannot end this section without mentioning the limitations of EC-TOPSIS:

- Firstly, the model integrates two methods for obtaining weights, without allowing the decision maker to choose to use other techniques, despite having an external input for a single solution. We believe this limitation could become a future opportunity for improvement, providing the decision-maker with other methods that can be chosen and combined in the model.

- Another point that could be explored is the randomization of the intervals for generating the criteria weights. We believe that we can add fuzzy set characteristics to the model.

- Finally, the definition of the final ranking, which we used the statistical measure MODE, but which can be applied to other measures or combinations of these to obtain improvements.

**6. Conclusions**

In the field of discussion on methods for obtaining criteria weights, experts are divided on the advantages and disadvantages of subjective and objective criteria weighting methods. There is a third way that integrates the two views of the problem, which gave rise to the integration of weighting methods and the emergence of hybrid methods. In these methods, the result of the weights generated is converted into a single set of weights, resulting in a single ranking. In these cases, integration is usually done using the product or geometric mean method.

In this article, we present EC-TOPSIS, which is a hybrid method that does not result in the design of a single set of weights but in a set of "t" iterations, which, when applied to TOPSIS, results in "t" sets of rankings. The innovation in this method, which sets it apart from others, consists firstly of creating a range of weights per criterion, with an upper and lower limit. The limits are obtained between the maximum and minimum values of each method used for each criterion. This implies that we use the characteristics of each method. The limits of each criterion form the interval for randomly generating the weights. Once "t" sets of weights have been generated, the phase of obtaining "t" sets of rankings begins. This process, which is integrated into the method, functions as a sensitivity analysis, where the decision maker will observe the variation in the ranking of the alternatives as a function of the set of weights used. From this set of data, the final ranking is obtained using the MODE statistical measure for the set of "t" positions occupied by each alternative. This modeling allows for the reduction of uncertainty in the construction of the final ranking, presenting a more consistent result for the decision-maker to consider.

The comparative analysis revealed a high degree of consistency among the methods, with EC-TOPSIS and CRITIC-COPRAS producing identical rankings (a1, a2, a4, a5, a6, and a3), while CRITIC-ARAS slightly differed by swapping the third and fourth positions (a1, a2, a5, a4, a6, and a3), underscoring methodological alignment with minor variations. Additionally, the ranking results derived from IDOCRIW and MEREC methods, when compared with the EC-TOPSIS rankings, reveal significant consistencies and differences.

Consistency in Top Rankings: Across all methods, a1 consistently emerges as the top-ranked alternative, indicating its robust performance regardless of the methodological approach. Similarly, a3 is consistently ranked the lowest, reflecting its weak comparative performance.

Middle-Tier Variations: While a2 and a4 maintain relatively strong positions in both EC-TOPSIS and the IDOCRIW/MEREC rankings, their exact rankings differ slightly depending on the methodology. EC-TOPSIS ranks a4 higher than a5, which aligns more closely with the IDOCRIW results than the MEREC ones, where the gap between these alternatives is less pronounced.

Methodological Sensitivity: The observed differences in middle-tier rankings highlight the sensitivity of ranking methodologies to the input criteria and their weighting schemes. EC-TOPSIS's emphasis on entropy-based measures may explain subtle discrepancies with IDOCRIW and MEREC, which integrate distinct decision-making perspectives.

In summary, the EC-TOPSIS rankings align closely with IDOCRIW and MEREC in identifying top and bottom alternatives but show slight divergences in ranking mid-performing alternatives, likely due to methodological differences. These findings underline the importance of method selection in MCDM processes for nuanced decision-making.

In future studies, a comprehensive analysis and comparison of weight calculation methods could be undertaken to establish robust criteria for evaluating their effectiveness. This would involve examining factors such as con-sistency, bias mitigation, the incorporation of expert knowledge, transparency, and flexibility. Such an analysis would systematically classify methods into "better" and "less effective" categories, offering valuable insights into their respective strengths and limitations.

## 7. Patents

The EC-TOPSIS Method has been certified as a Registered Computer Programme under registration number BR512024004431-0 at the National Institute of Industrial Property of the Ministry of Development, Industry, Commerce and Service of the Federative Republic of Brazil.

**Supplementary Materials:**

Computer Program Registration Certificate

**Author Contributions:**

Conceptualization, M.P.B.; methodology, M.P.B.; software, V.P..; validation, B.A. and S.A.; formal analysis, F.Y..; investigation, F.Y.; resources, M.P.B.; data curation, F.Y.; writing—original draft preparation. M.P.B., B.A., and F.Y; writing—review and editing, M.P.B. and F.Y.; visualization, M.P.B.; supervision, V.P.; project administration, M.P.B.; funding acquisition, M.P.B. All authors have read and agreed to the published version of the manuscript.

**Funding:**

This research received no external funding.

**Data Availability Statement:**

At the following link: https://pypi.org/project/ec-topsis/ the EC-TOPSIS tool in Python is available for readers to use free of charge.

**Conflicts of Interest:**

The authors declare no conflict of interest.

## Abbreviations

The following abbreviations are used in this manuscript:

| | |
|---|---|
| AHP | Analytic Hierarchy Process |
| ANP | Analytical Network Process |
| ARAS | Additive Ratio ASsessment |
| BWM | Best-Worst Method |
| CILOS | Criterion Impact Loss |
| COPRAS | Complex Proportional Assessment |
| CRITIC | Criteria Importance Through Intercriteria Correlation |
| NDM | Normalized Decision Matrix |
| EC-PROMETHEE | Entropy-CRITIC-PROMETHEE |
| EC-TOPSIS | Entropy-CRITIC-TOPSIS |
| EDAS | Evaluation based on Distance from Average Solution |
| ELECTRE | ÉLimination et Choix Traduisant la REalité (French) |
| FUCOM | Full Consistency Method |
| IDOCRIW | Integrated Determination of Objective CRIteria Weights |
| MARCOS | Measurement of Alternatives and Ranking according to COmpromise Solution |
| MCDA | Multi-Criteria Decision Analysis |
| MCDM | Multi-Criteria Decision-Making |
| MEREC | Method Based on the Removal Effects of Criteria |
| PROMETHEE | Preference Ranking Organization Method for Enrichment of Evaluation |
| SAPEVO-M | Simple Aggregation of Preferences Expressed by Ordinal Vectors—Multi-Decision Makers |
| SWARA | Step-wise Weight Assessment Ratio Analysis |
| TOPSIS | Technique for Order of Preference by Similarity to Ideal Solution |
| VIKOR | VlseKriterijumska Optimizacija I Kompromisno Resenje (Serbian) |